\documentclass[preprint,12pt]{elsarticle}
\usepackage{graphicx}
\usepackage{amssymb}
\usepackage{amsthm}
\usepackage{amsmath}

\begin{document}

\begin{frontmatter}

\title{Spatiotemporal coupled-mode theory in dispersive media under a dynamic modulation}

\author{Brenda Dana\footnote[1]{Corresponding author: E-mail address: brendada@post.tau.ac.il (Brenda Dana)}, Lilya Lobachinsky and Alon Bahabad \footnote[2]{Corresponding author: E-mail address: alonb@eng.tau.ac.il (Alon Bahabad); fax:+97236423508}}

\address{Department of Physical Electronics, School of Electrical Engineering, Fleischman Faculty of Engineering, Tel-Aviv University, Tel-Aviv 69978, Israel}

\begin{abstract}
A simple and general formalism for mode coupling by a spatial, temporal or spatiotemporal perturbation in dispersive materials is developed. This formalism can be used for studying various linear and non-linear optical interactions involving a dynamic modulation of the interaction parameters such as non-reciprocal phenomena, time reversal of signals and spatiotemporal quasi phase matching.\end{abstract}

\begin{keyword}
optical diffraction \sep spatiotemporal perturbation \sep quasi phase matching \sep coupled mode 
\MSC 30 \sep code 30.060 \sep 30.010
\end{keyword}

\end{frontmatter}

\section{Introduction}

Mode coupling is ubiquitous in physics, where different modes belonging to the same system or to different systems, can exchange energy through a coupling perturbation. There are many different formalisms of coupled-mode-theory in optics, describing a plethora of cases \cite{1077767,1075416,1074291,1073397,saleh1,yariv2007photonics}. The overwhelming majority of these formalisms relates to a spatial perturbation\footnote[3]{We exclude cases such as temporal perturbation of a light field to couple energy eigenmodes of an atom or a molecule \cite{Milonni_LaserPhysics2010}, as the modes themselves in this case are not propagating optical modes.}. In recent years several works involving temporal or spatiotemporal coupling between optical modes were published. In this case, instead of the usual spatial perturbation (e.g. grating), a dynamic modulation is induced on one or more of the interaction parameters. It is important to distinguish between such a dynamic modulation to the dynamics already present by virtue of the oscillating optical fields. An analogue can be made with the regular spatial case, where the spatial modulation (such as a diffraction grating or a Bragg mirror) is distinguished from the spatial modulation embodied by the wavelength of the optical fields. Works which used such a dynamic modulation were concerned with nonreciprocal propagation \cite{Kang_OptoAcousitcIsolator_Nphoton2011,Lira_ElectricallyDrivenNonReciprocity_PRL2012,Yu_Optical_Isolation_Nphoton2009}, with time reversal of optical signals  \cite{Sivan_TimeReversalPeriodicSystem_PRL2011,Yanik2004,Longhi2007} and with phase matching of optical frequency conversion processes \cite{Bahabad_STQPM_Nphoton2010}. In those works coupled mode equations were used implicitly or explicitly, while dispersion or phase mismatch were not always taken into account. In 1990 Yariv has given the result for the specific case of two modes coupled by a travelling wave induced by the electro-optic effect \cite{1077767,Zheng2006323,yariv1990optical} while recently, a formalism for spatiotemporal mode coupling in a system of parallel waveguides with no dispersion was derived \cite{Shteeman_TimeDependentCoupledModeTheoryWaveguides_JOSAB2010}.These cases imply that to date there is no general formalism treating \emph (dynamic modulation) for spatiotemporal mode coupling in general dispersive media. Here we develop such a formalism, which is general in the dispersive properties of the medium, number of modes and the geometry of the periodic spatiotemporal perturbation. This formalism can be applied to diverse cases, such as coupling between guided modes, coupling between free propagating and guided modes, and under non-depletion approximation also to nonlinear wave mixing and high harmonic generation. 

\section{Formalism}
\subsection{General spatiotemporal coupled mode analysis}
We start with the wave equation for the electric field   $\vec{E}(\vec{r},t)$  in a non-magnetic medium \cite{saleh1}:

\begin{equation}\label{eqn:1}
\nabla^2 \vec{E}(\vec{r},t)-\frac{1}{{c_0}^2}\frac{\partial^2}{\partial t^2}\vec{E}(\vec{r},t)=\mu_0\frac{\partial^2}{\partial t^2}\vec{P}(\vec{r},t)
\end{equation}

\noindent
where $c_0=\frac{1}{\sqrt{\mu_0\varepsilon_0}}$ is the speed of light with $\varepsilon_0$ and $\mu_0$  being the vacuum permittivity and vacuum permeability respectively and $\vec{P}(\vec{r},t)$ is the material polarization vector.\\
The polarization model we analyse assumes a complete separation between a linear transverse dispersive term $\chi^{(1)}(\vec{r}_{\bot},t)$ and a non-dispersive spatiotemporal perturbation term $\Delta\varepsilon(\vec{r},t)$. Here $\vec{r}_{\bot}=(x,y)$ and $\vec{r}=(x,y,z)$. For simplicity we also assume that both these terms are scalar. \emph{Unlike in regular spatial couple mode theory in which  $ \Delta\varepsilon=\Delta\varepsilon(\vec{r})$, the time dependence of $\Delta\varepsilon(\vec{r},t)$ in the present model allows to include a dynamic modulation as a coupling agent between modes. }Under these assumptions the polarization vector assumes the form \cite{Feng}:

\begin{flalign}\label{eqn:2}
&\vec{P}(\vec{r},t)=\vec{P}_L(\vec{r}_{\bot},t)+\Delta\vec{P}(\vec{r},t) = &\nonumber \\ &[\varepsilon_0\chi^{(1)}(\vec{r}_{\bot},t)*+\Delta\varepsilon(\vec{r},t)\cdot]\vec{E}(\vec{r},t)
\end{flalign}

\noindent
where $*$ stands for convolution. Substituting the polarization term given by Eq. \eqref{eqn:2} in Eq. \eqref{eqn:1}
and applying a Fourier transform in the time domain, results in:

\begin{flalign}\label{eqn:3}
&\nabla^2 \hat{\vec{E}}(\vec{r},\omega)+k^2(\vec{r}_{\bot},\omega)\hat{\vec{E}}(\vec{r},\omega)=&\nonumber \\& -\mu_0\omega^2[\Delta\hat{\varepsilon}(\vec{r},\omega)*\hat{\vec{E}}(\vec{r},\omega)]
\end{flalign}

\noindent
where $\hat{\vec{E}}(\vec{r},\omega)$ stands for the Fourier transform of $\vec{E}(\vec{r},t)$ and:

\begin{equation}\label{eqn:4}
k^2(\vec{r}_{\bot},\omega)=[1+\hat{\chi}^{(1)}(\vec{r}_{\bot},\omega)]\frac{\omega^2}{{c_0}^2}
\end{equation}

\noindent
In analogy with the case of having standard spatial perturbation \cite{yariv2007photonics}, for the solution of the wave equation in the case of having spatiotemporal perturbation we invoke a spatiotemporal variation of constants in which the modes amplitudes are both spatially and temporally dependent:  
\begin{equation}\label{eqn:5}
\vec{E}(\vec{r},t)= \sum_{m}^{}a_m(z,t)\vec{E}_m(\vec{r}_{\bot})e^{- i\beta_mz +i\omega_mt}
\end{equation}
\noindent
Here $\vec{E}_m(\vec{r}_{\bot})e^{- i\beta_mz +i\omega_mt}$ are the normal modes of the system satisfying the unperturbed Helmholtz wave equation given by \cite{yariv2007photonics}:
\begin{equation}\label{eqn:6}
[\nabla^2_{\bot}+k^2(\vec{r}_{\bot},\omega_m)-\beta_m^2]\vec{E}_m(\vec{r}_{\bot})=0
\end{equation}
\noindent
In the frequency domain Eq. \eqref{eqn:5} can be written as:
\begin{equation}\label{eqn:7}
\hat{\vec{E}}(\vec{r},\omega)= \sum_{m}^{}\hat{a}(z,\omega-\omega_m)\vec{E}_m(\vec{r}_{\bot})e^{- i\beta_mz}
\end{equation}
\noindent
Assuming the coupling dynamic perturbation is periodic in both space and time, we can represent it with a Fourier series expansion:
\begin{equation}\label{eqn:8}
\Delta\varepsilon(\vec{r}_{\bot},t)=\sum_{p,q}^{}\varepsilon_{p,q}(\vec{r}_{\bot})e^{i q \Omega t-i p Kz} \quad   p,q \in \mathbb{Z}
\end{equation}
\noindent
Where $\Omega$ and $K$ are the temporal and spatial fundamental frequencies of the modulation and the Fourier coefficients are given by:

\begin{flalign}\label{eqn:9}
&\varepsilon_{p,q}(\vec{r}_{\bot})=& \nonumber \\ &\frac{K\Omega}{4\pi^2}\int_{-\pi/K}^{\pi/K}dz\int_{-\pi/\Omega}^{\pi/\Omega}dt\Delta\varepsilon(\vec{r}_{\bot},z,t)e^{i q \Omega t-i p Kz}
\end{flalign}

\noindent
In the frequency domain the perturbation is written as:
\begin{flalign}\label{eqn:10}
&\Delta\hat{\varepsilon}(\vec{r}_{\bot},\omega)=&\nonumber \\ & \sum_{p,q}^{}\hat{\varepsilon}_{p,q}(\vec{r}_{\bot})e^{-i p Kz}\delta(\omega-q\Omega) \quad \quad p,q \in \mathbb{Z} 
\end{flalign}
\noindent
Let us examine the expression:
\begin{equation}\label{eqn:11}
k^2(\vec{r}_{\bot},\omega)\hat{\vec{E}}(\vec{r},\omega)
\end{equation}
\noindent
which appears in Eq. \eqref{eqn:3}. Using Eq. \eqref{eqn:7} and assuming that the envelopes $\hat{a}(z,\omega-\omega_m)$ are narrow-band relative to their respective carrier frequencies $\omega_m$ such that they are spectrally isolated from each other we can expand Eq. \eqref{eqn:11} in the dispersion terms as:

\begin{equation}\label{eqn:12}
\sum_{ m}^{ }\left[k^2(\vec{r}_{\bot},\omega_m)+\hat D(\omega)\right]\hat{a}(z,\omega-\omega_m)\vec{E}_m(\vec{r}_{\bot})e^{-i \beta_m z}
\end{equation}
\noindent
where $\hat D(\omega)=\sum_{l=1}^{\infty} \frac{1}{l!}\frac{\partial^l k^2(\vec{r}_{\bot},\omega) }{\partial\omega^l}|_{\omega=\omega_m}(\omega-\omega_m)^l$ is the dispersion operator in the frequency domain.

\noindent
We further make a functional approximation in which the dispersion terms are spatially independent. This way the spatial dependence of $ k^2(\vec{r}_{\bot},\omega) $ accounts for the guiding properties of the system, while the dispersive terms are defined within a volume of interest (such as the core of a wave-guide). Thus, $\frac{\partial^l}{\partial\omega^l}k^2(\vec{r}_{\bot},\omega)|_{\omega=\omega_m}\cong \frac{\partial^l}{\partial\omega^l}k^2(\omega)|_{\omega=\omega_m}$ .\\
\noindent
Substituting Eqs. (\ref{eqn:6}-\ref{eqn:7}), \eqref{eqn:10} and Eq. \eqref{eqn:11} into Eq. \eqref{eqn:3}, while using a spatially slowly varying envelope approximation  (SVEA) $\left|\frac{\partial^2}{\partial z^2}a_m(z,t)\right| \ll \left|\beta_m\frac{\partial}{\partial z}a_m(z,t)\right|$, we get:\\
\begin{flalign}\label{eqn:13}
&\sum_{m}^{}\vec{E}_m(\vec{r}_{\bot})e^{- i\beta_mz}\left[-2i\beta_m\frac{\partial}{\partial z}\hat{a}_m(z,\omega-\omega_m)+
\hat D(\omega)
\hat{a}_m(z,\omega-\omega_m)\right] &\nonumber \\ & =
-\mu_0\omega^2\sum_{p,q}^{}\hat{\varepsilon}_{p,q}(\vec{r}_{\bot})e^{-i pKz}\delta(\omega-\omega_n)*& \nonumber \\ & \sum_{n}^{}\vec{E}_n(\vec{r}_{\bot})\hat{a}_n(z,\omega-\omega_n)e^{-i \beta_nz}
\end{flalign}
\noindent
We use the usual normalization for a 1 [W] power flow in the propagation direction which results in the orthogonality relation \cite{yariv2007photonics}:
\begin{flalign}\label{eqn:14}
&<\vec{E}_m(\vec{r}_{\bot})|\vec{E}_n(\vec{r}_{\bot})> \nonumber \\ &=\int d\vec{r}_{\bot} \vec{E}_m(\vec{r}_{\bot})^*\vec{E}_n(\vec{r}_{\bot})=\frac{2 \omega_m \mu}{|\beta_m|}\delta_{mn}
\end{flalign}
\noindent
where $<\cdot|\cdot>$ stands for inner product and $\delta_{mn}$ is either the Kronecker delta function for bound modes or the Dirac delta function for radiation modes. With this orthogonality condition we project Eq. \eqref{eqn:13} on a specific mode $<m|$ to get:

\begin{flalign}\label{eqn:15}
&\frac{2 \omega_m}{|\beta_m|} e^{- i\beta_mz}\left[-2i\beta_m\frac{\partial}{\partial z}\hat{a}_m(z,\omega-\omega_m)+ 
\hat D(\omega)
\hat{a}_m(z,\omega-\omega_m)\right] =& \nonumber\\& 
-\omega^2\sum_n\sum_{p,q}C_{mn}^{pq}e^{-i(\beta_n+pK)z}
\hat{a}_n(z,\omega-\omega_n)*\delta(\omega-q\Omega)
\end{flalign}
\noindent
with the coupling coefficient:
\begin{equation}\label{eqn:16}
C_{mn}^{pq}=\int d\vec{r}_{\bot} \vec{E}_m(\vec{r}_{\bot})^*\hat{\varepsilon}_{p,q}(\vec{r}_{\bot})\vec{E}_n(\vec{r}_{\bot})
\end{equation}
\noindent
Transforming back to the time domain Eq. \eqref{eqn:15} takes the form:
\begin{align}\label{eqn:17}
&\left[ \frac{\partial}{\partial z}-\frac{1}{2i\beta_m}
D(t)
\right]a_m(z,t)\nonumber\\
&=\frac{|\beta_m|}{4i\omega_m\beta_m} \sum_n \sum_{p,q} C_{mn}^{pq} e^{-i(\beta_n-\beta_m+pK)z} e^{-i\omega_mt}\frac{\partial^2}{\partial t^2} a_n(z,t)e^{i(\omega_n+q\Omega)t}
\end{align}
\noindent
where $D(t)=\sum_{l=1}^{\infty}\frac{(-i)^l}{l!}\frac{\partial^lk^2(\omega) }{\partial\omega^l}|_{\omega=\omega_m}\frac{\partial^l}{\partial t^l}=\sum_{l=1}^{\infty}\frac{2k(\omega_m)(-i)^l}{l!}(\frac{\partial^l k(\omega) }{\partial\omega^l}|_{\omega=\omega_m})\frac{\partial^l}{\partial t^l}$ is the dispersion operator in the time domain \cite{McDonald}.

\emph{Eq. \eqref{eqn:17} is the major result of this paper. Unlike in regular spatial perturbation mode coupling, this result includes the effect of a periodic applied dynamic modulation through the appearance of the $q\Omega$ temporal frequencies. }

\noindent
If we now further assume a second-order \emph{temporal} SVEA: 
$\left| \frac{\partial^2}{\partial t^2}a_n(z,t) \right| \ll \left| \omega_n\frac{\partial}{\partial t}a_n(z,t)\right| \ll \left|\omega_n^2 a_n(z,t)\right|$
and that $| q\Omega |$ is much smaller than $\omega_n$ we are left with:

\begin{flalign}\label{eqn:18}
&\left[ \frac{\partial}{\partial z}-\frac{1}{2i\beta_m}
D(t)
\right]a_m(z,t)&\nonumber\\
&=\kappa_m \sum_n \sum_{p,q} C_{mn}^{pq} a_n(z,t)e^{i\Delta \omega_{m,n,q}t}e^{-i\Delta k_{m,n,q}z}
\end{flalign}
\noindent
with $\kappa_m=\frac{i|\beta_m|(\omega_n+q\Omega)^2}{4\omega_m\beta_m}$ and where $\Delta k_{m,n,q}=\beta_n-\beta_m+pK$ is the momentum phase mismatch and $\Delta \omega_{m,n,q}=\omega_n-\omega_m+q\Omega$ is the energy phase mismatch. This form is appealing as the spatial and temporal phase mismatch terms appear with the same functional form. The appearance of the temporal phase mismatch is the result of allowing for a dynamic modulation. This term does not appear in regular - spatial coupled mode theory.

\subsection{The special case of two coupled modes under first order dispersion approximation}
Eq. \eqref{eqn:18}  describes a general case of mode coupling caused by a temporal and/or spatial perturbations in a dispersive media. In many cases of interest there are only two coupled modes (there might be other modes but the coupling with these other modes is negligible) and only the first dispersion term is important. Let the two coupled mode be designated with the indices $1$ and $2$. We assume that the coupling is made through a specific $(p,q)$ order of the perturbation. In this case the coupled mode equations reduce to:
\begin{align}\label{eqn:19}
&\left( \frac{\partial}{\partial z}+\frac{k(\omega_1)}{\beta_1}\frac{1}{v_g(\omega_1)}\frac{\partial}{\partial t}\right)a_1(z,t)\nonumber\\
&=\kappa_1 C_{12}^{p,q} a_2(z,t)e^{i \Delta \omega t}e^{-i\Delta k z}
\end{align}

\begin{align}\label{eqn:20}
&\left( \frac{\partial}{\partial z}+\frac{k(\omega_2)}{\beta_2}\frac{1}{v_g(\omega_2)}\frac{\partial}{\partial t}\right)a_2(z,t)\nonumber\\&=\kappa_2({C_{12}^{p,q}})^*a_1(z,t)e^{-i \Delta \omega t}e^{i\Delta k z}
\end{align}
\noindent
with the phase mismatch components $\Delta k=\beta_2-\beta_1+pK$ and $\Delta \omega=\omega_2-\omega_1+q\Omega$. We have used the relation $C_{21}^{-p,-q}=({C_{12}^{p,q}})^*$ which is easily derived from Eq. \eqref{eqn:16}. In addition we used $\frac{\partial}{\partial\omega}k^2(\omega)|_{\omega=\omega_m}= 2\frac{k(\omega_m)}{v_g(\omega_m)}$ with the usual definition of the group velocity ${v_g}^{-1}(\omega)= \frac{\partial}{\partial\omega}k(\omega)$.

We would like to comment that these last two equations can be further reduced to known cases in the literature. With no temporal modulation:
$\Delta\omega=0$ and $q\Omega=0$, they reduce reduce to regular spatial coupled mode equations \cite{1077767,1073397} .  
For the case of nonlinear frequency conversion under the non-depletion approximation: With temporal modulation: Eq. (\ref{eqn:20}) is virtually identical in form to Eq. (2) in Ref.\cite{Bahabad_STQPM_Nphoton2010}; with no temporal modulation this same equation is reduced to the form found in  Eq. (2.7.11) in Ref.\cite{boyd2003nonlinear}.  

\section{Discussion}
In this work we developed a general coupled-mode formalism under the presence of a dynamic modulation leading to a temporal or a spatiotemporal perturbative coupling of modes. \emph{This dynamic modulation is absent from regular spatial optical coupled mode theory. } This generalization can not only describe known phenomena such as optical non-reciprocity and time reversal but can also leads to new ones - such as the formation of a breathing soliton pair with no nonlinearity \cite{Time-dependent coupled mode analysis of parallel waveguides}.
  
Our model is not restricted to any specific system or mechanism for inducing the perturbation. It applies to any number of modes, as well as to a continuum of modes. It includes high order dispersion terms. The model requires that the coupling perturbation can be approximated as non-dispersive, compared to the linear susceptibility term, as is explicit in Eq.\eqref{eqn:2}. This condition is satisfied for example when the perturbation is applied through the electrooptic effect. This can be proved using Miller's rule \cite{boyd2003nonlinear} applied to the electrooptic nonlinear polarization term. 
Another implicit required condition is that the spatial perturbation and temporal modulations are much slower than those of the optical frequencies of the relevant modes. Otherwise the problem becomes one of effective medium approximation.  

\section{References}
\bibliographystyle{unsrt}

\begin{thebibliography}{10}
%

\bibitem{1077767}
A.~Yariv.
\newblock Coupled-mode theory for guided-wave optics.
\newblock {\em Quantum Electronics, IEEE Journal of}, 9(9):919--933, Sep 1973.

\bibitem{1075416}
H.A. Haus, W-P Huang, S.~Kawakami, and N.~Whitaker.
\newblock Coupled-mode theory of optical waveguides.
\newblock {\em Lightwave Technology, Journal of}, 5(1):16--23, Jan 1987.

\bibitem{1074291}
Amos Hardy and William Streifer.
\newblock Coupled mode theory of parallel waveguides.
\newblock {\em Lightwave Technology, Journal of}, 3(5):1135--1146, Oct 1985.

\bibitem{1073397}
Shun-Lien Chuang.
\newblock Application of the strongly coupled-mode theory to integrated optical
  devices.
\newblock {\em Quantum Electronics, IEEE Journal of}, 23(5):499--509, May 1987.

\bibitem{saleh1}
B.E.A. Saleh and M.C. Teich.
\newblock {\em Fundamentals of Photonics}.
\newblock Wiley Series in Pure and Applied Optics. John Wiley \& Sons, 2007.

\bibitem{yariv2007photonics}
A.~Yariv and P.~Yeh.
\newblock {\em Photonics: Optical Electronics in Modern Communication}.
\newblock The Oxford Series in Electrical and Computer Engineering Series.
  Oxford University Press, 2007.

\bibitem{Milonni_LaserPhysics2010}
P.W. Milonni and J.H. Eberly.
\newblock {\em Laser Physics}.
\newblock Wiley, 2010.

\bibitem{Kang_OptoAcousitcIsolator_Nphoton2011}
MS~Kang, A~Butsch, and P~St~J Russell.
\newblock Reconfigurable light-driven opto-acoustic isolators in photonic
  crystal fibre.
\newblock {\em Nature Photonics}, 5(9):549--553, 2011.

\bibitem{Lira_ElectricallyDrivenNonReciprocity_PRL2012}
H.~Lira, Z.~Yu, S.~Fan, and M.~Lipson.
\newblock Electrically driven nonreciprocity induced by interband photonic
  transition on a silicon chip.
\newblock {\em Physical Review Letters}, 109(3):33901, 2012.

\bibitem{Yu_Optical_Isolation_Nphoton2009}
Z.~Yu and S.~Fan.
\newblock Complete optical isolation created by indirect interband photonic
  transitions.
\newblock {\em Nature photonics}, 3(2):91--94, 2009.

\bibitem{Sivan_TimeReversalPeriodicSystem_PRL2011}
Yonatan Sivan and John~B Pendry.
\newblock Time reversal in dynamically tuned zero-gap periodic systems.
\newblock {\em Physical Review Letters}, 106(19):193902, 2011.

\bibitem{Yanik2004}
Mehmet~Fatih Yanik and Shanhui Fan.
\newblock Time reversal of light with linear optics and modulators.
\newblock {\em Phys. Rev. Lett.}, 93:173903, Oct 2004.

\bibitem{Longhi2007}
Stefano Longhi.
\newblock Stopping and time reversal of light in dynamic photonic structures
  via bloch oscillations.
\newblock {\em Phys. Rev. E}, 75:026606, Feb 2007.

\bibitem{Bahabad_STQPM_Nphoton2010}
A.~Bahabad, M.M. Murnane, and H.C. Kapteyn.
\newblock {Quasi-phase-matching of momentum and energy in nonlinear optical
  processes}.
\newblock {\em Nature Photonics}, 4(8):571--575, 2010.

\bibitem{Zheng2006323}
Guoliang Zheng and Weilong She.
\newblock Generalized wave coupling theory of linear electro-optic effect in
  absorbent medium.
\newblock {\em Optics Communications}, 268(2):323 -- 329, 2006.

\bibitem{yariv1990optical}
A.~Yariv.
\newblock {\em Optical Electronics}.
\newblock The Oxford Series in Electrical and Computer Engineering Series.
  Oxford University Press, 1990.

\bibitem{Shteeman_TimeDependentCoupledModeTheoryWaveguides_JOSAB2010}
Vladislav~R Shteeman, Inna Nusinsky, and Amos~A Hardy.
\newblock Time-dependent coupled mode analysis of parallel waveguides.
\newblock {\em JOSA B}, 27(4):735--741, 2010.

\bibitem{Feng}
Tian Feng, Y.-Z. Wu, and P.-D. Ye.
\newblock Improved coupled-mode theory for anisotropic waveguide modulators.
\newblock {\em Quantum Electronics, IEEE Journal of}, 24(3):531--536, 1988.

\bibitem{McDonald}
G.S. McDonald, J.M. Christian, and T.F. Hodgkinson.
\newblock Optical soliton pulses with relativistic characteristics.
\newblock pages 4--6, 2010.


\bibitem{Time-dependent coupled mode analysis of parallel waveguides}
B.~Dana, B.~Malomed and A.~Bahabad.
\newblock {Breathing solitary-pulse pairs in a linearly coupled system}
\newblock {To be published in Optics Letters}.


\bibitem{boyd2003nonlinear}
R.W. Boyd.
\newblock {\em Nonlinear Optics}.
\newblock Electronics \& Electrical. Acad. Press, 2003.


\end{thebibliography}

\end{document}